\documentstyle{article}
\title{\vspace*{-2.1cm}\hspace*{8.7cm} {\large gr-qc/9309024,
CGPG-93/9-3}\vspace*{.9cm}\\
{\bf Reality Conditions in Nonperturbative Quantum Cosmology}}
\author{\\Guillermo A. Mena Marug\'an\vspace*{.6cm}\\
Center for Gravitational Physics and Geometry,\\ The Pennsylvania State
University, 104 Davey Laboratory,\\ University Park, PA 16802-6300, USA.
\vspace*{.4cm}\\ On leave from: {\it Instituto de Matem\'aticas y F\'{\i}sica
Fundamental},\\ {\it C.S.I.C., Serrano 121, 28006 Madrid, Spain.}\\ }
\date{September, 1993}

\begin{document}
\renewcommand{\thefootnote}{\fnsymbol{footnote}}

\maketitle
\large
\setlength{\baselineskip}{.825cm}

\begin{center}
{\bf Abstrat}
\end{center}

We carry out the nonperturbative canonical quantization of several
types of cosmological models that have already been studied in the
geometrodynamic formulation using the complex path-integral approach. We
establish a relation between the choices of complex contours in the path
integral and the sets of reality conditions for which the metric
representation is well defined, proving that the ambiguity in the selection
of complex contours disappears when one imposes suitable reality
conditions. In most of the cases, the wave functions defined by means of
the path integral turn out to be non-normalizable and cannot be accepted as
proper quantum states. Moreover, the wave functions of the Universe picked
out in quantum cosmology by the no-boundary condition and the tunneling
proposals do not belong, in general, to the Hilbert space of quantum
states. Finally, we show that different sets of reality conditions can lead
to equivalent quantum theories. This fact enables us to extract physical
predictions corresponding to Lorentzian gravity from quantum theories
constructed with other than Lorentzian reality conditions.

\newpage

\section {Introduction}

The nonperturbative canonical quantization programme proposed by Ashtekar
[1,2] is accepted nowadays as one of the most promising approaches to
construct a consistent theory of quantum gravity. In spite of the success
of this programme, developed systematically over the last seven years
[2,3], the nonperturbative quantization of the full theory of general
relativity remains still incomplete. In order to gain insight into the
kind of difficulties involved in quantizing gravitational systems, an
increasing number of works in the literature have been devoted to the
nonperturbative quantization\footnote{In the following, we will understand
``nonperturbative quantization'' to refer to the result of the
nonperturbative canonical quantization programme proposed by Ashtekar
[2].} of truncated models in gravity [4,5]. Apart from providing a good
arena to test the applicability of the quantization procedure, the
completion of this nonperturbative quantization in minisuperspaces of
cosmological interest is clearly relevant inasmuch as it allows us to
extract physical predictions in quantum cosmology [5].

A particularly appealing possibility consists in attempting the
nonperturbative quantization of minisuperspace models that have been
previously studied in quantum cosmology using the standard techniques of
the geometrodynamic formulation [6]. In this way, one could check the
validity of the results obtained in quantum cosmology and
reach a better physical interpretation of the mathematical framework
employed in the nonperturbative quantization, eventually adopting
proposals from the geometrodynamic formulation to solve current problems in
nonperturbative quantum gravity.

With these motivations in mind, we will carry out in this work the
nonperturbative quantization of two families of minisuperspace models
that have been considered in the literature as examples of
exactly soluble systems in the complex path-integral formalism of
quantum cosmology.

First, we will study a class of homogeneous and
spherically symmetric models provided with a minimally coupled massless
scalar field with exponential potential.  Models of this kind appear, for
instance, in the dimensional reduction of five-dimensional Kaluza-Klein
gravity with cosmological constant [7,8], and their classical solutions
display, in general, either exponential or power-law inflation [8,9,10].
The path-integral analysis  of these models was discussed in detail in
Ref. [10].

The second type of models that we will analyze are a family of homogeneous
and anisotropic minisuperspaces with cosmological constant, that
contains the locally rotationally symmetric (LRS) Bianchi types I and III and
the Kantowski-Sachs model as particular cases. The complex path integral
between fixed metric configurations was thoroughly studied for these
minisuperspaces in Ref. [11].

For all the above models, we will implement to completion the
nonperturbative canonical quantization programme [2]. We briefly recall
that, in order to apply this quantization procedure to a system with
constraints, one first selects a complete set of complex functions on
phase space that is closed under Poisson brackets [2,3]. This set is
promoted to a $\star$-algebra of elementary operators, capturing the
complex conjugation relations between classical variables as
$\star$-relations [2,12]. The $\star$-algebra of elementary operators is
then represented on a chosen vector space, with the quantum states
provided by the kernel of all the operators that represent the classical
constraints of the system. Finally, one can fix the inner product in the
space of quantum states by imposing the $\star$-relations between
elementary operators as adjointness conditions [13]. These adjointness
requirements are usually called reality conditions [2].

Reality conditions seem to play a double role in the nonperturbative canonical
quantization programme: they simultaneously select the quantum theory
associated with a specific section of the complex phase space of the
system (the section chosen by reality conditions when
considered as complex conjugation relations [2,12]) and determine the inner
product in that theory by demanding some adjointness requirements.

There is a certain parallel between the need to consider complex
functions on phase space in the nonperturbative quantization,
and the definition of the Euclidean path integral in quantum cosmology as
an integral over complex fields and metrics. In the nonperturbative
quantization, reality conditions remove the ambiguity that was introduced
by the complexification of the phase space. In path-integral cosmology
[14,15], on the other hand, the complex contours of integration are
severely restricted by the requirement of convergence when one imposes the
reality of the metric and matter fields on the boundaries of the manifold
where one is integrating [16]. Nevertheless, the ambiguity in this latter
case is only partially removed, and more than one choice of inequivalent
complex contours of integration are usually acceptable (even if one tries
to demand additional conditions for the path integral [16]).

Using that, for the models considered in this work, both the
nonperturbative quantization and the path integral between fixed
geometrodynamic configurations can be carried out successfully, we will
show that the possible choices of complex contours in the geometrodynamic
path integral correspond in fact to the selection of different reality
conditions for the system. In particular, once one has fixed an adequate
set of reality conditions, the complex contours of integration
turn out to be essentially unique.

We will also explore another related topic that at present remains unclear
in the nonperturbative programme: the relation between the quantum
theories obtained by demanding different sets of reality conditions.  In
general, one of the main technical difficulties encountered in the
implementation of the nonperturbative quantization scheme is the
imposition of the reality conditions associated with Lorentzian gravity
(specially in Ashtekar variables [12]). A tentative way out of this
problem could consist of completing the nonperturbative quantization by
employing simpler reality conditions, and recovering somehow the
Lorentzian predictions from the quantum theory so constructed. For the
models studied in this paper, we will prove that sets of non-Lorentzian
reality conditions can lead in some cases to quantum theories that are
equivalent to that corresponding to Lorentzian gravity, so that one can
actually extract from them all the relevant physics.

\section {Minisuperspaces: Nonperturbative Canonical\newline Quantization}

{\sl {\bf 2.1. Isotropic models with scalar field}}

\vspace*{.4cm}
The first class of minisuperspaces that we want to consider are a family
of homogeneous and isotropic models with a minimally coupled scalar field.
We will restrict the metric of these models to be of the form
\begin{equation} ds^2=-N^2
\frac{dt^2}{a^2(t)}+a^2(t)\;d\Omega_3^2\;,\end{equation}
and the scalar field to be homogeneous and with a potential of exponential
type:
\begin{equation} V(\phi)=\alpha\;\cosh(2\phi)+\beta\;\sinh(2\phi)
\;.\end{equation}
Here, $d\Omega_3^2$ is the metric on the unit three-sphere, $\alpha$
and $\beta$ are real constants, and $N$, $a$ and $\phi$ denote,
respectively, the numerical values of the lapse function, the scale
factor and the scalar field  (in the system of units chosen in Ref.
[10]).

In the new variables
\begin{equation} x=a^2\;\cosh (2\phi)\;,\;\;\;\;\;y=a^2\sinh(2\phi)
\;,\end{equation}
the Hamiltonian constraint of these minisuperspaces takes the simple
expression
\begin{equation} {\cal H}=\frac{1}{2}\left(-4p_x^2+4p_y^2+\alpha x+
\beta y-1\right)=0\;,\end{equation}
where $p_x$ and $p_y$ are the momenta canonically conjugate to $x$ and $y$.

We notice that the transformation (3) maps the ``physical'' region $a,\phi
\in I\!\!\!\,R$ into the light cone of the origin in the real $xy$ space.
However, instead of restricting the domain of the variables $x$ and $y$,
we will regard them as our elementary complex configuration variables,
following, in that sense, the same quantization approach that was adopted
in Ref. [10].

On the other hand, all models with $\alpha\neq\beta$ can be related to the
case $\beta=0$ by a linear transformation of coordinates [10]. It will
then suffice to concentrate our attention on two types of scalar field
potential: those corresponding to $\beta=0$ and to $\alpha=\beta$. We will
require in addition that $\alpha>0$, so that the potential for the scalar
field is at least bounded from below. Finally, we will employ from now on the
notation
\begin{equation} k=\frac{1}{\alpha}\;;\end{equation}
so, $k>0$ for all the models under consideration.

\vspace*{.4cm}
{\sl {\bf 2.1.1. The cosh potential}}

\vspace*{.4cm}

Let us analize first the case $\beta=0$, $\alpha=k^{-1}>0$. For this model,
it is easy to check that the variables
\renewcommand{\theequation}{\arabic{equation}.a}
\begin{equation}Q=y+8\,k\,p_xp_y\;,\;\;\;\;\;\;\;\;\;\;\;\;\;P=p_y\;\;\;\;
\end{equation}
form a pair of canonically conjugate observables, i.e., of functions on phase
space that commute with the Hamiltonian constraint. Introducing then the
additional coordinates
\setcounter{equation}{5}
\renewcommand{\theequation}{\arabic{equation}.b}
\begin{equation}
H=x+4\,k\,(p_y^2-p_x^2)\;,\;\;\;\;\;\;T=p_x\;,\end{equation}
\renewcommand{\theequation}{\arabic{equation}}
we obtain a complete set of (complex) phase-space variables. Equations (6)
define a canonical transformation of coordinates that is analytic and
invertible everywhere. This transformation is generated by the function
\begin{equation} F=x\,T+y\,P+4\,k\,P^2T-\frac{4}{3}k\,T^3\;.\end{equation}

The variable $H$ is essentially equal to the Hamiltonian of the system
\begin{equation} {\cal H}\propto H-k=0\;.\end{equation}
The coordinate $T$, on the other hand, plays the role of an intrinsic
time, since $T$ and $H$ are canonically conjugate to each other.

Given the simplicity of the Hamiltonian constraint and the Poisson brackets
in the new coordinates (6), it is now almost straightforward to achieve the
nonperturbative quantization of this minisuperspace. We choose $Q$, $P$, $H$
and $T$ as our set of basic functions on phase space, and represent the
corresponding algebra of elementary operators on the space of distributions
in the complex variables $T$ and $P$ (this choice of representation will be
very convenient for our later discussion in\linebreak section 3 ).

In the adopted representation (and taking $\hbar=1$ from now on), the
action of the elementary operators can be defined in the following way:
\begin{equation} \hat{P}\Phi=P\Phi(T,P),\;\;\;\hat{Q}\Phi=i
\frac{\partial\Phi}{\partial P}(T,P),\;\;\;\hat{T}\Phi=T\Phi(T,P),\;\;\;
\hat{H}\Phi=i\frac{\partial\Phi}{\partial T}(T,P),\end{equation}
and the solutions to the Hamiltonian constraint are simply given by
the expression
\begin{equation} \Phi(T,P)=e^{-ikT}\,f(P)\;.\end{equation}

In order to complete the nonperturbative quantization, we have to impose a set
of reality conditions, and find the associated inner product.
For Lorentzian gravity, the variables $Q,$ $P$, $H$ and
$T$ are real, and thus we conclude that
\begin{equation} \hat{Q}^{\star}=\hat{Q}\;,\;\;\;\hat{P}^{\star}=\hat{P}\;,
\;\;\;\hat{H}^{\star}=\hat{H}\;,\;\;\;\hat{T}^{\star}=\hat{T}\;.\end{equation}
Among these relations, only the two first can be promoted to
self-adjointness conditions in the quantum theory, because $\hat{H}$
and $\hat{T}$ are not observables and, therefore, their action in the space
of quantum states is not well defined. Nevertheless, requiring that $\hat{Q}$
and $\hat{P}$ be self-adjoint will be enough to fix the inner product,
since these variables form a complete set of observables for the model
[3,13]. On the other hand, if we insist on the connection between complex
conjugation and $\star$-relations, the conditions on $\hat{H}$ and $\hat{T}$
in (11) allow us to restrict the domain of the variable $T$ to be the real
axis. This restriction is consistent with the Lorentzian dynamics, for
$\dot{T}=N\alpha\{H,T\}=N\alpha$ is always real if $N\in I\!\!\!\,R$.

To determine the inner product, we first notice that the dependence of the
quantum states on $T$ is entirely fixed (each state is completely
characterized by the distribution $f(P)$). As a consequence, the
inner product can be made time independent by choosing an adequate
$T$-dependence in the integration measure. In this way, one arrives at
a product of the form
\begin{equation} <\Psi|\Phi>=\frac{i}{2}\int dP\wedge d\overline{P}
\mu(P,\overline{P})\overline{h(P)}f(P)\;,\end{equation}
where $\Psi=\exp(-ikT)h(P)$, the symbol $^{^{-\!\!-}}$ denotes
complex conjugation  and the measure $\mu(P,\overline{P})$
must be such that $\hat{Q}$ and $\hat{P}$ are self-adjoint. After a short
calculation, we conclude that
\begin{equation} <\Psi|\Phi>=\int_{I\!\!\!\,R} dP\; \overline{h}(P) f(P)\;,
\end{equation}
so that the Hilbert space of quantum states is $L^2(I\!\!\!\,R)$.

Before continuing our discussion, we would like to make a few remarks
on the quantum theory that we have constructed. We have seen above that,
for Lorentzian gravity, the variables $x$ and $y$ in (3) should be not
only real, but restricted to the light cone of the origin. However, we
have carried out the quantization without imposing the analogue of this
constraint in the new variables (6). In this sense, the quantum theory
that we have obtained contains contributions from metrics and scalar
fields that are not purely Lorentzian. One reason that justifies this
approach is simply that the classical evolution associated to the
Hamiltonian (4) does not leave invariant the light cone of the origin in
the real $xy$ plane; thus, if we take $x$ and $y$ as the basic
configuration variables of the model, we cannot consistently restrict
their domain to the proposed region. On the other hand, our main interest
lies in the comparison between the results of the nonperturbative
quantization and the complex path-integral formalism. Since the
path-integral analysis of this model was performed without restricting the
domain of the variables $x$ and $y$ [10], we will adopt here the same kind
of strategy in order to maintain the parallel in the quantization as
far as possible.

Let us study now more general sets of reality conditions than those
corresponding to Lorentzian gravity. For our purposes in section 3, where
we will explore the relation between complex contours for the path
integral and reality conditions, it will suffice to consider conditions of
the type
\begin{equation} \hat{T}^{\star}=e^{-i2\varphi}\hat{T}\;,\;\;\;
\hat{H}^{\star}=e^{i2\varphi}\hat{H}+k(1-e^{i2\varphi})\;,
\end{equation}
\begin{equation} \hat{Q}^{\star}=e^{-i2\theta}\hat{Q}\;,\;\;\;\hat{P}^{\star}
=e^{i2\theta}\hat{P}-i\epsilon(1+e^{i2\theta})\;,\end{equation}
where $\epsilon\!\in I\!\!\!\,R$ and $\varphi,\theta\in[0,\pi)$ are three
real constants. Note that the $\star$-relations (14,15) are compatible with
the commutation relations, provided that
\begin{equation}(a\hat{X}+b\hat{Y})^{\star}=\overline{a}\hat{X}^{\star}+
\overline{b}\hat{Y}^{\star}\;,\;\;\;\;(\hat{X}\hat{Y})^{\star}=\hat{Y}
^{\star}\hat{X}^{\star}\end{equation}
for any complex constants $a$ and $b$ and any operators $\hat{X}$ and
$\hat{Y}$ in the
$\star$-algebra. The $\star$-relation for $\hat{H}$ has been chosen in
such a way that the operator version of the Hamiltonian constraint (8)
and its $\star$-analogue are equivalent. In the $(T,P)$ representation,
conditions (14) can be interpreted as restricting the domain of the
variable $T$ to lie in the real axis rotated by an angle $\varphi$:
\begin{equation} T\in\Gamma\equiv\{e^{i\varphi}\tau, \;\,\tau\in I\!\!\!\,R
\}\;,\;\;\;\varphi\in[0,\pi)\;.\end{equation}
This is clearly compatible with the Hamiltonian dynamics for
non-Lorentzian times obtained by a similar rotation of the lapse function
($N=\exp(i\varphi)\,c$ with $c\in I\!\!\!\,R$), since $\dot{T}=N\alpha$.

Following similar arguments to those presented above for the Lorentzian
reality conditions, it is not difficult to conclude that the inner product
determined by conditions (15), imposed as self-adjointness relations in
the operators $\hat{Q}$ and $\hat{P}$, takes the simple expression
\begin{equation} <\Psi|\Phi>=e^{i\theta}\int_{\gamma}dP
\;\overline{h(P)} f(P)\;,\end{equation}
where $\gamma$ is an infinite line in the complex $P$ plane that can be
parametrized as
\begin{equation} \gamma\equiv\{e^{-i\theta}p+i\epsilon\,,\;\,p\in I\!\!\!\,R
\}\;,\end{equation}
and $\theta$ and $\epsilon$ are the same constants that appear in equation
(15). From now on, we will denote the corresponding Hilbert space as
$L^2(\gamma)$.

\vspace*{.4cm}
{\sl {\bf 2.1.2. The exponential potential}}

\vspace*{.4cm}
We turn now to the case of a purely exponential potential for the scalar
field: $\alpha=\beta>0$. For this model, the following transformation of
phase-space coordinates provides a new set of canonical variables:
\renewcommand{\theequation}{\arabic{equation}.a}
\begin{equation}
Q=y+4\,k\,p_y^2-k\,\left(p_y-p_x\right)^2,\;\;\;\;\;P=p_y-p_x\;,\end{equation}
\setcounter{equation}{19}
\renewcommand{\theequation}{\arabic{equation}.b}
\begin{equation}H=x+y+4\,k\,\left(p_y^2-p_x^2\right),\;\;\;\;\;\;\;T=p_x\;.
\;\;\;\;\;\;\end{equation}
\renewcommand{\theequation}{\arabic{equation}}
This canonical transformation is generated by the function
\begin{equation} F=\frac{P}{2}(y-x-H)-\frac{1}{16kP}(x+y-H)^2\;.\end{equation}
We note that the transformation (20) is analytic and invertible
everywhere, even though the generator (21) is singular at $P=0$ (this
singularity will nevertheless have some consequences in the $xy$
representation, as we will see in the following section).

Just as in the $\beta=0$ model, the variables $Q$ and $P$ are a canonically
conjugate pair of observables, $H$ is essentially the Hamiltonian, which is
given again by equation (8), and $T$ corresponds to an intrinsic time. To
quantize the model, we choose the space of distributions in the variables
$P$ and $H$ as our representation space, and introduce the following action
for the elementary operators
\begin{equation}\hat{P}\Phi\!=\!P\Phi(H,P),\;\,\hat{Q}\Phi\!=i
\frac{\partial\Phi}{\partial Q}(H,P),\;\,\hat{H}\Phi\!=\!H\Phi(H,P),\;\,
\hat{T}\Phi\!=\!-i\frac{\partial\Phi}{\partial H}(H,P),\end{equation}
with the derivatives defined in the distributional sense.
In this representation, the general solution to the Hamiltonian constraint
adopts the compact expression
\begin{equation} \Phi(H,p)=\delta(H-k)\,f(P)\;,\end{equation}
where $\delta$ is the Dirac function and $f(P)$ any distribution.

To fix the inner product, we still have to impose an admissible set of
reality conditions. We will first concentrate our attention on reality
conditions of the type (14,15). The inner product can be clearly made $H$
independent, because all quantum states (23) are characterized by their
dependence on $P$. The set of reality conditions (15), imposed as
adjointness relations among the observables of the theory, determine then
a unique inner product of the form (18). In particular, the inner product for
Lorentzian gravity (corresponding to conditions (11)) is given again by
equation (13).

Another possibility that we want to investigate is that of quantizing the
theory assuming a restricted domain for the variable $P$.
This quantization approach is consistent in principle, since the
classical evolution leaves invariant any domain of definition for $P$.
We are particularly interested (for our later discussion in section 3)
in studying those cases in which $P$ is restricted to a
half-infinite segment with endpoint at $P=0$:
\begin{equation} P\in\gamma_+\equiv\{e^{-i\theta}\,p,\;\,p\in I\!\!\!\,R^+\}
\;,\end{equation}
with $\theta\in [0,2\pi)$ a fixed angle. For this kind of models, the
reduced phase space can be taken to be the cotangent bundle over $\gamma_+$.
An adequate set of elementary variables in that reduced phase space
is provided by $P$ and the new coordinate
\begin{equation} q=PQ\;,\end{equation}
where $Q$ is the momentum canonically conjugate to $P$. The choice of $q$
as the generalized momentum variable is motivated by the fact that its
associated vector field, $P(\partial/\partial P)$, is complete on
$\gamma_+$, while the vector field that correspond to $Q$, $(\partial/
\partial P)$, fails to satisfy this condition [5].

To quantize the system, we choose $(H,T,P,q)$ as our set of elementary
variables, which is obviously closed under the Poisson-bracket structure.
We then adopt the same $(H,P)$ representation that was selected before
in this subsection, except that we substitute now the operator $\hat{Q}$
in (22) by the new elementary operator
\begin{equation} \hat{q}\Phi=iP\frac{\partial\Phi}{\partial P}(H,P)\;.
\end{equation}
The general solution to the Hamiltonian constraint is still given by
equation (23). On the other hand, we impose as reality conditions that,
in addition to relations (14),
\begin{equation} \hat{q}^{\star}=\hat{q}\;,\;\;\;\;\hat{P}^{\star}=
e^{i2\theta} \hat{P}\;.\end{equation}
The second of these conditions is the $\star$-analogue of the complex
conjugation relation derived from (24). The first condition in
(27) has been chosen to guarantee that the $\star$-relations (that must
satisfy the requirements (16)) are compatible with the commutator
$[\hat{q},\hat{P}]=i\hat{P}$.

Generalizing our previous analysis of the implementation of the reality
conditions, it is straightforward to arrive at an inner product of
the form
\begin{equation} <\Psi|\Phi>=\int_{\gamma_+}\frac{dP}{P}\;\overline{h(P)}
f(P)\;,\end{equation}
with $\gamma_+$ the half-infinite contour defined in (24), and $\Psi=
\delta(H-k) h(P)$. The Hilbert space of quantum states is thus
$L^2(\gamma_+,P^{-1}dP)$.

We finally notice that the quantum theory that we have constructed is in fact
unitarily equivalent to that with Hilbert space equal to $L^2(\gamma_+)$
(the Hilbert space with inner product given by (18) evaluated at
$\gamma=\gamma_+$) and operator $\hat{q}$ defined as
\begin{equation} \hat{q}\tilde{\Phi}=i\left(P\frac{\partial \;\;}{\partial P}
+\frac{1}{2}\right)\tilde{\Phi}(H,P)\;.\end{equation}
The isomorphism between the Hilbert spaces $L^2(\gamma_+,P^{-1}dP)$
and $L^2(\gamma_+)$ is provided by
\begin{equation}
\Phi(H,P)=\delta(H-k)f(P)\;\stackrel{I}{\longrightarrow}\;\tilde{\Phi}(H,P)=
\delta(H-k)\frac{f(P)}{\sqrt{P}},\end{equation}
\[f(P)\in L^2(\gamma_+,P^{-1}dP)\;;\;\;\;\;\;\;\;\;\;\;
\;\tilde{f}(P)=\frac{f(P)}{\sqrt{P}}\in L^2(\gamma_+)\;.\]
The action of the operator $\hat{q}$ in (29) corresponds, on the other
hand, to the symmetrized product of the operators $\hat{Q}$ and $\hat{P}$, as
given by equation (22). In this sense, the quantum theory associated with
reality conditions (14) and (27) is equivalent to that with elementary
operators $(\hat{H},\hat{T},\hat{P},\hat{Q})$ defined through (22) and
Hilbert space $L^2(\gamma_+)$. We will use this alternative quantization in
the next section for our discussion of the connection between choices of
complex contours for the path integral and sets of reality conditions.

\vspace*{.4cm}
{\sl {\bf 2.2. Anisotropic models}}

\vspace*{.4cm}
In addition to the previous models, we want to analyze also a class of
anisotropic minisuperspaces with cosmological constant whose spacetime
metric can be expressed in the generic form
\begin{equation} ds^2=-N^2\frac{dt^2}{a^2(t)}+a^2(t)\,dr^2+b^2(t)\,
d\Omega_2^2\;.\end{equation}
Here, $N$, $a$ and $b$ are the numerical values of the lapse function
and the two scale factors of the model (in adequate units, see Ref. [11]),
$d\Omega_2^2$ denotes the metric on a compact orientable two-manifold
of constant scalar curvature equal to $2k$, and $k=+1$, 0 or -1 [11].
Depending on the value of $k$, the metric (31) describes either the
Kantowski-Sachs model ($k=1$), the LRS Bianchi type I ($k=0$) or the
LRS Bianchi type III ($k=-1$).

After defining the new variable
\begin{equation} c=a^2\,b\;,\end{equation}
the Hamiltonian constraint of these systems can be written
\begin{equation} {\cal H}= \frac{1}{2}(-4 p_c p_b+\lambda
b^2-k)=0\;,\end{equation}
where $\lambda$ is the rescaled cosmological constant [11], and $p_c$
and $p_b$ are the momenta canonically conjugate to $c$ and $b$.

Paralleling the analysis of the isotropic models with a scalar field,
we introduce the following canonical transformation of variables
\renewcommand{\theequation}{\arabic{equation}.a}
\begin{equation} Q=c+\frac{\lambda b^3-6b\,p_bp_c}{6p_c^2}\;,\;\;\;\;\,
P=p_c\;,\;\end{equation}
\setcounter{equation}{33}
\renewcommand{\theequation}{\arabic{equation}.b}
\begin{equation} H=-4p_cp_b+\lambda
b^2\;,\;\;\;\;\;\;\;\;\,\,T=\frac{b}{4p_c}\;. \end{equation}
\renewcommand{\theequation}{\arabic{equation}}
A generating function for the above transformation is given by
\begin{equation} F=c\,P+\frac{\lambda b^3-3b\,H}{12P}\;.\end{equation}

Once again, the variables $Q$ and $P$ provide a conjugate pair of observables,
the Hamiltonian constraint adopts the simple expression (8) and $T$ plays
the role of an intrinsic time. The only relevant difference with respect
to the minisuperspaces studied before is the existence of a singularity in
the canonical transformation (34) at $P=p_c=0$. We will return to
this point at the end of this section.

To achieve the nonperturbative canonical quantization of these models, we
choose $(Q,P,H,T)$ as our set of elementary complex variables,
take the space of distributions in $H$ and $P$ as our representation
space, and define the action of the elementary operators as in equation
(22). The general solution to the Hamiltonian constraint is then of the
form (23), with the parameter $k$ equal to either 1, 0 or $-1$.

The same line of reasoning that was presented above for the scalar field
model with exponential potential leads us to conclude that, for reality
conditions of the type (14,15), or for conditions (14) and (27) if the
range of the variable $P$ is restricted to lie on a half-infinite contour,
the inner product is unique and given by formula (18) or (28),
respectively. Therefore, the Hilbert space of quantum states is again
equal to either $L^2(\gamma)$ or $L^2(\gamma_+)$ (assuming that we adopt
for both classes of reality conditions the representation defined through
equation (22)).

The only caveat with respect to this result is the existence of a
singularity at the origin of $P$ in the canonical transformation (34).
This singularity may affect the quantum theories with Hilbert spaces of
the form $L^2(\gamma(\epsilon=0))$, where $\gamma(\epsilon=0)$ (as given
by (19)) is a contour obtained by rotating the real axis around the
origin, and therefore contains the point $P=0$\footnote{Note that all the
contours with $\theta=\pi/2$ in (19) are equivalent to
$\gamma(\epsilon=0,\theta=\pi/2)$.}. Nevertheless, we notice that one can
always avoid getting contributions from the singularity at the origin of
$P$ by requiring that $f(P=0)=0$ for all quantum states $f$. This
condition can be consistently imposed on any $f\in
L^2(\gamma(\epsilon=0))$, because a function in the Hilbert space
$L^2(\gamma)$ is defined only almost everywhere in $\gamma$.

\section {Complex Path Integral and Reality Conditions}

In the previous section, we have completed the nonperturbative
quantization of the considered minisuperspace models by using a set of
canonically conjugate variables that mix the geometrodynamic configuration
and momenta coordinates. We now want to analyze the relation between the
representation employed in this nonperturbative quantization and the metric
representation\footnote{From now on, we will use the term ``metric
representation'' to designate any representation in which the metric and
the matter fields of the model act as multiplicative operators.},
discussing thereafter the validity of the complex path-integral approach
in geometrodynamics. In particular, we will show that the selection of
different complex contours in the path integral correspond in fact to the
choice of different sets of reality conditions for the nonperturbative
quantization.

\vspace*{.4cm}
{\sl {\bf 3.1. Scalar field with cosh potential}}

\vspace*{.4cm}
Let us study first the case of the homogeneous and spherically symmetric
model provided with a scalar field with hyperbolic cosine potential. It
is not difficult to check that, for this minisuperspace, the change from
the representation ($T,P$), defined through equation (9), to the
metric representation ($x,y$) used in Ref. [10] is obtained by means of the
transformation
\begin{equation} f(x,y)=\int_{\gamma}dP\int_{\Gamma}dT\,e^{iF(x,y,T,P)}
\Phi(T,P)\;,\end{equation}
where $F$ is the generating function that appears in (7), $\Phi(T,P)$
are the quantum states (10), and $\Gamma$ and $\gamma$ are, respectively,
the contours (17) and (19) selected by the set of reality conditions (14,15).
The restriction to these contours in (36) is due to the fact that, in
principle, the states of the quantum theory constructed with reality
conditions (14,15) are defined only for $T\in\Gamma$ and $P\in\gamma$.

The transformation (36) guarantees that, if all the boundary terms that come
from integration by parts vanish, the action of the operators
\begin{equation}\hat{x}=\hat{H}+4k(\hat{T}^2-\hat{P}^2),\;\;\,\;\hat{p}_x=
\hat{T},\;\;\;\,\hat{y}=\hat{Q}-8k\hat{T}\hat{P},\;\;\;\,\hat{p}_y=\hat{P}
\end{equation}
(with $(\hat{Q},\hat{P},\hat{H},\hat{T})$ defined in (9)) is given in
the $(x,y)$ representation by the standard expressions
\begin{equation}\hat{x}f=x\,f(x,y),\;\;\;\hat{p}_xf=-i\frac{\partial f}
{\partial x} (x,y),\;\;\;\hat{y}f=y\,f(x,y),\;\;\;\hat{p}_yf=-i\frac
{\partial f}{\partial y}(x,y)\;.\end{equation}

In the following, we will concentrate our attention only on real values of
$x$ and $y$. This is motivated by our final intention of
studying the results of the complex path-integral formalism in the
geometrodynamic formulation, results that were obtained in Ref. [10] by
assuming that $x$ and $y$ are real on the boundaries of the manifold.

The $T$-integration that appears in (36):
\begin{equation}I(x,P)=\int_{\Gamma} dT\,e^{ixT+i4kP^2T-i\frac{4}{3}kT^3-ikT}
\end{equation}
can be defined in a convergent way $\forall x\in I\!\!\!\,R$ only for three
choices of the contour $\Gamma$:
\renewcommand{\theequation}{\arabic{equation}.a}
\begin{equation} \Gamma_1\equiv I\!\!\!\,R \;\;\;\;\;\;\,\; \rightarrow
\;\;\;I(x,P)\propto A_i(X(x,P^2))\;,
\;\;\;\;\;\;\;\;\;\;\;\;\;\;\;\;\;\;\;\;\;\;\;\;\;\;\;\;\,
\end{equation}
\setcounter{equation}{39}
\renewcommand{\theequation}{\arabic{equation}.b}
\begin{equation} \Gamma_2\equiv e^{i\frac{\pi}{3}} I\!\!\!\,R \;\;\;
\rightarrow
\;\;\;I(x,P)\propto [A_i(X(x,P^2))+iB_i(X(x,P^2))]\;,\end{equation}
\setcounter{equation}{39}
\renewcommand{\theequation}{\arabic{equation}.c}
\begin{equation} \Gamma_3\equiv e^{i\frac{2\pi}{3}} I\!\!\!\,R\; \;
\rightarrow \;\;\;I(x,P)\propto [A_i(X(x,P^2))-iB_i(X(x,P^2))]\;,
\end{equation}
\renewcommand{\theequation}{\arabic{equation}}
where $A_i$ and $B_i$ are the Airy functions [17], and
\begin{equation} X(x,P^2)=\left(1-4P^2-\frac{x}{k}\right)
\left(\frac{k}{2}\right)^{\frac{2}{3}} \;.\end{equation}
The contours $\Gamma_j$ (with $j=1,2,3$) in (40) are selected by the
subclass of reality conditions (14)
\begin{equation} \hat{T}^{\star}=e^{-i\frac{2\pi}{3}(j-1)}\hat{T}\;,\;\;
\hat{H}^{\star}=e^{i\frac{2\pi}{3}(j-1)}\hat{H}+k\left(1-e^{i\frac{2\pi}{3}
(j-1)}\right)\;.\end{equation}
For the rest of contours $\Gamma$ in (17), the representation $(x,y)$ is
not well defined in the whole real $x$ axis.

Let us consider now the integration in $P$:
\begin{equation} f(x,y)=\int_{\gamma}dP\,e^{iyP}f(P)I(x,P)\;.\end{equation}
In order to obtain the wanted change of representation, it is necessary
that the above integral be convergent $\forall y\in I\!\!\!\,R$ and $\forall
f\in L^2(\gamma)$. Taking into account the asymptotic behaviour of the Airy
functions [17] and the expressions (40.b,c) and (41), it is
possible to show that, for the contours $\Gamma_2$ and $\Gamma_3$,
this is indeed the case if $\gamma$ is either equal to the real axis
or given by equation (19) with $\theta\in (2\pi/3,\pi)$ when
$\Gamma=\Gamma_2$, or with $\theta\in (0,\pi/3)$ when $\Gamma=\Gamma_3$.
We recall that the reality conditions (15) that pick out $\gamma=
I\!\!\!\,R$ are
\renewcommand{\theequation}{\arabic{equation}.a}
\begin{equation} \gamma= I\!\!\!\,R\;\;\;\;\;\rightarrow\;\;\hat{Q}^
{\star}=\hat{Q}\;,\;\;\hat{P}^{\star}=\hat{P}\;.\end{equation}
\setcounter{equation}{43}
For the contour $\Gamma_1$, and $I(x,P)$ provided by (40.a), a similar
analysis leads to the conclusion that there are two types of acceptable
contours of integration $\gamma$ of the form (19):
a) $\gamma= I\!\!\!\,R$, and b) $\gamma=\gamma(\theta,\epsilon)$ with
$\theta\in (\pi/3,2\pi/3)$ and $\epsilon$ any real constant. In
particular, $\gamma=-i I\!\!\!\,R$ is an admissible choice of contour,
that corresponds to the reality conditions
\renewcommand{\theequation}{\arabic{equation}.b}
\begin{equation} \gamma=-i I\!\!\!\,R\;\;\rightarrow\;\;\hat{Q}^{\star}=
-\hat{Q}\;,\;\;\hat{P}^{\star}=-\hat{P}\;.\end{equation}
\renewcommand{\theequation}{\arabic{equation}}
The contours $\Gamma_1=\gamma=I\!\!\!\,R$, on the other hand, are those
selected by the reality conditions (11) for Lorentzian gravity.
Notice that, for $\gamma=I\!\!\!\,R$ and a given function $f(P)\in L^2(I\!
\!\!\,R)$, the three functions $f(x,y)$ obtained in (43) for the different
contours $\Gamma_j$ ($j=1,2,3$) are linearly dependent. This is due to the
fact that the linear combination of contours
$\Gamma=\Gamma_1-\Gamma_2+\Gamma_3$ can be distorted to zero in the
integral (39).

For all the above choices of contours $\Gamma$ and $\gamma$, the change of
representation defined by (36) can be inverted to recover the initial
$(T,P)$ representation. The result can be written with the compact notation
\begin{equation} f(P)=e^{ikT}\Phi(T,P)=e^{ikT-i4kP^2T+i\frac{4}{3}kT^3}
\int_{\overline{\Gamma}}dx e^{-ixT}\int_{\overline{\gamma}}dy e^{-iyP}
f(x,y)\;,\end{equation}
where $\overline{\Gamma}$ and $\overline{\gamma}$ are the contours obtained
from $\Gamma$ and $\gamma$, respectively, by taking complex
conjugation\footnote{In fact, for all the contours $(\Gamma,\gamma)$
considered here, the functions $f(x,y)$ given by (36) can
always be defined for $x\in \overline{\Gamma}$ and $y\in \overline{\gamma}$,
and the integrals appearing in (45) are convergent.}.
{}From equations (45) and (18) it is now straightforward to arrive at the
expression
of the inner product in the $(x,y)$ representation. In particular, for
reality conditions corresponding to Lorentzian gravity (i.e., for $\Gamma=
\gamma=I\!\!\!\,R$), the inner product turns out to be
\begin{equation} <f|g>=\int_{I\!\!\!\,R}dP\int_{I\!\!\!\,R}dx\int_{I\!\!\!
\,R}dy\int_{I\!\!\!\,R}dx'\int_{I\!\!\!\,R}dy' e^{i(x-x')T+i(y-y')P}
\;\overline{f}(x,y)g(x',y')\;.\end{equation}
It is possible to check that this inner product is in fact $T$ independent
when evaluated at quantum solutions to the Hamiltonian constraint (4) (for
$\beta=0$). On the other hand, equation (46) can be further simplified
when the order of integration is interchangeable.

We turn now to compare the results of the complex path-integral formalism
and the nonperturbative quantization approach. The propagation amplitudes
between final and initial real geometrodynamic configurations were calculated
in Ref. [10] by computing the path integrals that provide these amplitudes
along suitable complex contours. It was shown there that there exist three
possible inequivalent choices of infinite complex contours for the
integration in the lapse function $N$, each of them leading to different
propagation amplitudes. The three contours selected in Ref.
[10], and their corresponding amplitudes, are
\renewcommand{\theequation}{\arabic{equation}.a}
\begin{equation}
N=i\,n-\tilde{\epsilon},\;n\in I\!\!\!\,R;\;\;\;\;\;\;
G=\int_{I\!\!\!\,R}dw\,
e^{w(y-y')} A_i(X(x,-w^2))A_i(X(x',-w^2))\;,\;\;\end{equation}
\setcounter{equation}{46}
\renewcommand {\theequation}{\arabic{equation}.b}
\begin{equation}N=i\,n+\tilde{\epsilon},\;n\in I\!\!\!\,R;\;\;\;
G=\int_{I\!\!\!\,R}dw\, e^{iw(y-y')}
[A_i(X)B_i(X')+A_i(X')B_i(X)]\;,\;\,\end{equation}
\setcounter{equation}{46}
\renewcommand{\theequation}{\arabic{equation}.c}
\begin{equation}N\in I\!\!\!\,R^+\cup i\,I\!\!\!\,R^+;\;\;\;\;\;\,
G=\int_{I\!\!\!\,R}dw\, e^{iw(y-y')}
[A_i(X)+iB_i(X)][A_i(X')+iB_i(X')]\;,\end{equation}
\renewcommand{\theequation}{\arabic{equation}}
\begin{equation} X=X(x,w^2)\;,\;\;\;\;X'=X(x',w^2)\;,\end{equation}
with $\tilde{\epsilon}>0$, $X$ defined in (41), and ($x,y$) and $(x',y')$,
respectively, the final and initial fixed values of $x$ and $y$.

The propagation amplitude in (47.a) can be seen then as a function of the
form (43), with $I(x,P)$ chosen as in (40.a) (i.e., $\Gamma=\Gamma_1$),
$\gamma=-i \,I\!\!\!\,R$ and
\renewcommand{\theequation}{\arabic{equation}.a}
\begin{equation} f(P)=f_1(P)\equiv e^{-iy'P} A_i(X(x',P^2))\;,\end{equation}
\setcounter{equation}{48}
$\!\!$up to numerical factors. Similarly, the amplitude (47.c) can be obtained
from equations (43) and (40.b) (i.e., $\Gamma=\Gamma_2$), with
$\gamma=I\!\!\!\,R$ and
\renewcommand{\theequation}{\arabic{equation}.b}
\begin{equation}f(P)=f_2(P)\equiv
e^{-iy'P}[A_i(X(x',P^2))+iB_i(X(x',P^2))]\;.\end{equation}
\setcounter{equation}{48}
Finally, a linear combination of the lapse function contours in (47.b) and
(47.c) (and, therefore, of their respective propagation amplitudes)
leads to a wave function of the form (43), for $I(x,P)$ given by (40.c)
($\Gamma=\Gamma_3$), $\gamma=I\!\!\!\,R$ and
\renewcommand{\theequation}{\arabic{equation}.c}
\begin{equation} f(P)=f_3(P)\equiv
e^{-iy'P}[A_i(X(x',P^2))-iB_i(X(x',P^2))]\;.\end{equation}
\renewcommand{\theequation}{\arabic{equation}}
For $f(P)=f_j(P)$ fixed for each contour $\Gamma=\Gamma_j$ ($j=1,\,2,\,3$),
the rest of contours $(\Gamma,\gamma)$ for which the real metric
representation (43) is well defined provide us with no new propagation
amplitude.

In this way, the three possible inequivalent choices of complex
path-integral contours lead to wave functions in three different quantum
theories, each of them obtained by imposing a different set of
reality conditions that allows the existence of a well defined real $(x,y)$
representation. In other words, the various admissible choices of complex
contours in the path integral correspond in fact to the selection of
different sets of reality conditions for the model. Therefore,
the ambiguity in the determination of the complex contours of integration
is removed entirely when one accepts, since the very beginning, a suitable
set of reality conditions.

Another question that we want to address is whether the wave functions
constructed from the complex path integrals describe admissible
quantum states, i.e., normalizable wave functions. The answer, in general,
is in the negative: even though the wave function (49.a) belongs to the
corresponding Hilbert space $L^2(\gamma)$ for $\gamma=-i\,I\!\!\!\,R$, it is
easy to check that the wave functions (49.b) and (49.c) are not square
integrable in $\gamma=I\!\!\!\,R$, and so are not quantum states in the
associated Hilbert space $L^2(I\!\!\!\,R)$.

Moreover, it was shown in Ref. [10] that, in the model under consideration,
the wave functions selected by the no-boundary proposal [18] are simply the
propagation amplitudes with fixed initial values $x'$ and $y'$ equal to
zero. Since these wave functions are a particular subclass of
those studied before, we conclude that, in general, the no-boundary
condition does not choose normalizable quantum states (at least in the
minisuperspace implementation discussed in [10,11]). Therefore, it seems
inconsistent to use this proposal to pick out a wave function of the
Universe in quantum cosmology.

\vspace*{.4cm}
{\sl {\bf 3.2. Scalar field with exponential potential and anisotropic
models}}

\vspace*{.4cm}
In this subsection, we will simultaneously analyze the
metric representation for both the anisotropic minisuperspaces and the
homogeneous model with scalar field and exponential potential. For these
two kinds of models, the change to the metric representation employed in
Refs. [10,11] can be obtained by means of a transformation of the form
\begin{equation} f(u,v)=\int_{\gamma} \frac{dP}{\sqrt{P}}\int_{C}dH\;
e^{iF(u,v,H,P)}\Phi(H,P)\;,\end{equation}
where $\Phi(H,P)$ are the quantum states (23), the contour $C$ of
integration for $H$ is selected by reality conditions (14), and $\gamma$
is the straight line (19) associated with conditions (15).
In equation (50), $(u,v)$ are the geometrodynamic configuration variables
$(x,y)$ for the scalar field model or $(c,b)$ for the anisotropic models,
and $F(u,v,H,P)$ denotes the generating function (21) or (35), respectively.

The factor $1/\sqrt{P}$ has been included in (50) to ensure that the
operators
\renewcommand{\theequation}{\arabic{equation}.a}
\begin{equation} \hat{u}=\hat{x}=\hat{H}-\hat{Q}-k(\hat{P}-4\hat{T}^2)
\;,\;\;\;\;\;\;\hat{p}_u=\hat{p}_x=\hat{T}\;,\;\;\;\;\,\end{equation}
\setcounter{equation}{50}
\renewcommand{\theequation}{\arabic{equation}.b}
\begin{equation} \hat{v}= \hat{y}=\hat{Q}-k[4(\hat{P}+\hat{T})^2-
\hat{P}^2]\;,\;\;\;\;\;\hat{p}_v=\hat{p}_y=\hat{P}+\hat{T}\end{equation}
\setcounter{equation}{50}
for the scalar field model, and
\renewcommand{\theequation}{\arabic{equation}.c}
\begin{equation}\hat{u}=\hat{c}=
\hat{Q}+\frac{16}{3}\lambda\hat{T}^3\hat{P}-4
\hat{P}^{-1}\left(\hat{H}\hat{T}-\frac{i}{2}\right),\;\;\;\;\hat{p}_u=
\hat{p}_c=\hat{P}\;,\end{equation}
\setcounter{equation}{50}
\renewcommand{\theequation}{\arabic{equation}.d}
\begin{equation} \hat{v}=\hat{b}=4\hat{T}\hat{P}\;,\;\;\;\;\;\;\;\;\;\;
\hat{p}_v=\hat{p}_b=4\lambda\hat{T}^2\hat{P}-\frac{1}{4}\hat{P}^{-1}
\hat{H}\end{equation}
\renewcommand{\theequation}{\arabic{equation}}
for the anisotropic models, act in the $(u,v)$ representation in the standard
way
\begin{equation} \hat{u}f=u\,f(u,v)\;,\;\;\hat{p}_uf=-i\frac{\partial f}
{\partial u}(u,v)\;,\;\;\hat{v}f=v\,f(u,v)\;,\;\;\hat{p}_vf=-i\frac
{\partial f}{\partial v}(u,v)\;,\end{equation}
assuming that all the boundary terms that come from integration by parts in
(50) can be disregarded. Notice that in (51.c) we have chosen the
symmetric ordering for the product $\hat{H}\hat{T}$ that appears in the
operator $\hat{u}$, and that we have supposed that the singularity of
$\hat{P}^{-1}$ at $P=0$ (see (51.c,d)) can be handled without problems
in our calculations\footnote{This can be achieved by requiring that
$f(P=0)=0$ $\forall f\in L^2(\gamma)$ whenever $P=0\in \gamma$.}.

Reality conditions (14) determine, for the variable $H$, infinite contours $C$
that always contain the point $H=k$. Substituting the expression (23) for
the quantum states $\Phi$, the $H$ integration in (50) can then be
straightforwardly performed. The result can be written with the compact
notation
\begin{equation} f(u,v)=f(w,z)=\int_{\gamma}\frac{dP}{\sqrt{P}}\,
e^{iwP-\frac{i}{4P}g(z)}f(P)\;.\end{equation}
For the scalar field model with exponential potential, $(w,z)$ are the
following linear combination of the configuration variables $(u,v)$
\renewcommand{\theequation}{\arabic{equation}.a}
\begin{equation} w=\frac{v-u}{2}+\frac{k}{2}=\frac{y-x}{2}+\frac{k}{2}
\;,\;\;\;\;\;\;\;z=\frac{u+v}{2}-\frac{k}{2}=\frac{x+y}{2}-\frac{k}{2}
\end{equation}
and $g(z)$ is given by
\setcounter{equation}{53}
\renewcommand{\theequation}{\arabic{equation}.b}
\begin{equation}g(z)=\frac{z^2}{k}\;,\end{equation}
\setcounter{equation}{53}
with $k$ strictly positive. For the anisotropic models, on the other hand,
\renewcommand{\theequation}{\arabic{equation}.c}
\begin{equation}
w=u=c\;,\;\;\;\;\;\;z=v=b\;,\;\;\;\;\;\;
g(z)=kz-\frac{\lambda}{3}z^3
\end{equation}
\renewcommand{\theequation}{\arabic{equation}}
and $k=+1$, 0 or $-1$.

Equation (53) provides the transformation to the metric representation for
real $w$ and $z$ if the integral on the right-hand side exists $\forall
f\in L^2(\gamma)$. It is not difficult to prove that this restricts the
possible contours $\gamma$ of the type (19) to be of the form
$\gamma=I\!\!\!\,R+i\,\epsilon$, with $\epsilon$ any real constant.

The case $\epsilon=0$ ($\gamma=I\!\!\!R$) deserves special comments.
In the variables $(w,z)$ introduced in (54), the Hamiltonian constraints of
the models studied in this subsection adopt the generic expression
\begin{equation} {\cal H}\propto 4\,k\,p_wp_z
+\frac{dg}{dz}(z)=0\;,\end{equation}
with $p_w$ and $p_z$ the conjugate momenta to $w$ and $z$, and $g(z)$ defined
in (54). The solutions to the quantum version of this constraint are of the
form $\exp{(iwP)}\psi(z)$, where $\psi(z)$ satisfies the Schr\"odinger
equation
\begin{equation} i\frac{d\psi}{dz}=\frac{1}{4 k P}\frac{dg}{dz}\;\psi
\;,\end{equation}
that is obviously ill defined if $P=0$. For $P\neq 0$, we arrive at
the family of wave functions \[ e^{iwP-\frac{i}{4P}g(z)}\;.\] The quantum
states (53) can then be interpreted as a superposition of such wave
functions, assuming that there exists no contribution with $P=0$. For
$\gamma=I\!\!\!\,R$ in (53) this last requirement can be fulfilled only if
$f(P=0)=0$, in agreement with our discussion of the anisotropic models in
section 2. We will hence impose that $f(P=0)=0$ for all functions $f\in
L^2(I\!\!\!\,R)$, both for the scalar field model with exponential potential
and for the anisotropic minisuperspaces.

Let us study now the transformation to the metric representation
for quantum theories constructed by restricting the domain of the
variable $P$ to a half-infinite segment with endpoint at $P=0$, and
with reality conditions of the type (14) and (27). The Hilbert spaces for
these theories are isomorphic to $L^2(\gamma_+)$, for contours $\gamma_+$
defined through equation (24). Paralleling the analysis
presented above for the Hilbert spaces $L^2(\gamma)$, it is possible to
prove that the transformation (53) is still valid in these cases if we
substitute the contour $\gamma$ by $\gamma_+$. The change of representation
(53) can be defined $\forall w,z\in I\!\!\!\,R$ and $\forall f\in
L^2(\gamma_+)$ if and only if $\gamma_+$ equals either the positive or the
negative real axis. For our purposes, it will suffice to analyze the case
$\gamma_+=I\!\!\!\,R^+$, because the Hilbert space $L^2(I\!\!\!\,R)$
(with the imposition $f(0)=0$ for all functions in it) can be considered as
the direct sum of $L^2(I\!\!\!\,R^+)$ and $L^2(I\!\!\!\,R^-)$.
The results for $L^2(I\!\!\!\,R^-)$ can thus be derived from those
corresponding to $L^2(I\!\!\!\,R)$ and $L^2(I\!\!\!\,R^+)$.

For all the choices of contours for which the real $(w,z)$ representation
is well defined, i.e., for $\gamma=I\!\!\!\,R+ i\,\epsilon$ or
$\gamma_+=I\!\!\!\,R^+$, the transformation (53) can be inverted to recover
the $(H,P)$ representation:
\begin{equation} \Phi(H,P)=\delta(H-k)f(P)=\delta(H-k)\sqrt{P}\,e^{-\frac{i}
{4P}g(z)}\int_{I\!\!\!\,R}dw\,e^{-iwP}f(w,z)\;,\end{equation}
where the variable $P$ runs over the specified contour $\gamma$ or $\gamma_+$.
It is possible to check that this expression is in fact $z$ independent when
$f(w,z)$ is a wave function of the form (53). Using this equation, one can
easily derive the formula for the inner product in the $(w,z)$
representation. For example, for reality conditions corresponding to
Lorentzian gravity (that select, we recall, the contour $\gamma=I\!\!\!\,R$),
the inner product results in being
\begin{equation} <f|g>=\int_{I\!\!\!\,R}dP\,|P|\int_{I\!\!\!\,R}dw
\int_{I\!\!\!\,R}dw'\,e^{i(w-w')P}\;\overline{f}(w,z)g(w',z)\;,\end{equation}
that again can be showed to be $z$ independent.

Let us relate now these results with those obtained in Ref. [10] and [11],
where the complex path integrals that provide the propagation amplitudes
between real geometrodynamic configurations were computed. The analysis in
those references shows that there exist three inequivalent choices of
infinite contours of integration in the lapse function for which the
propagation amplitudes are well defined:
\begin{equation} N\in\gamma_1=i\,I\!\!\!\,R+\tilde{\epsilon}\;,\;\;\;
N\in\gamma_2=i\,I\!\!\!\,R-\tilde{\epsilon}\;,\;\;\;N\in\gamma_3=i\,
I\!\!\!\,R-\{0\}\;,\end{equation}
and $\tilde{\epsilon}>0$ a constant. Employing the notation introduced in
(54), the propagation amplitudes between final and initial
configurations $(w,z)$ and $(w',z')$, respectively, can be rewritten as
[10,11]
\begin{equation} G=\int_{\gamma_j}\frac{dN}{N}e^{N(w-w')+\frac{1}{4N}
[g(z)-g(z')]}\;,\end{equation}
with $\gamma_j$ ($j=1,2,3$) one of the three contours given by (59). In
these cases, the integral (60) can be calculated exactly [11]. Note, on
the other hand, that all the contours $\gamma_1$ (or $\gamma_2$)
obtained with different constants $\tilde{\epsilon}>0$ lead to the
same propagation amplitudes $G$.

There exists another inequivalent choice of integration contour such that
the propagation amplitude (60) converges, namely, the positive
imaginary axis [11],
\begin{equation} N\in\gamma_4=i\,I\!\!\!\,R^+\;.\end{equation}
For this half-infinite contour, the amplitude (60) provides us
with a Green function for the minisuperspace Wheeler-DeWitt equation [14],
i.e., with a solution to the Hamiltonian constraint except when the final
and initial configurations coincide.

With the change of coordinates $N=iP$, the propagation amplitude (60)
becomes a wave function of the form (53), for
\begin{equation}
f(P)=\frac{1}{\sqrt{P}}e^{-iw'P+\frac{i}{4P}g(z')} \end{equation}
and
\renewcommand{\theequation}{\arabic{equation}.a}
\begin{equation} P\in
I\!\!\!\,R-i\,\tilde{\epsilon},\;\,\tilde{\epsilon}>0\;\;\;\;
{\rm if} \;\;N\in\gamma_1\;,\;\;\;\;\;\;\;\;\;\;\;\;\;\;\;\;\;\;
\end{equation}
\setcounter{equation}{62}
\renewcommand{\theequation}{\arabic{equation}.b}
\begin{equation}P\in I\!\!\!\,R+i\,\tilde{\epsilon},\;\,
\tilde{\epsilon}>0\;\;\;\;{\rm if}\;\;N\in \gamma_2\;,\;\;\;\;\;\;
\;\;\;\;\;\;\;\;\;\;\;\;\end{equation}
\setcounter{equation}{62}
\renewcommand{\theequation}{\arabic{equation}.c}
\begin{equation} \;\;\;\;\;\;\;\;\;\;\;\;\;\;\;\;\;\;\;\;\;\;\;P\in
I\!\!\!\,R\;\;\;\;\;\;\;\;\;\;\;\;\;\;\;\;\;\;\;\;\;\;\,
{\rm if} N\in \gamma_3 \;\;\;\;({\rm and}\;\;f(P=0)=0)\;,\end{equation}
\setcounter{equation}{62}
\renewcommand{\theequation}{\arabic{equation}.d}
\begin{equation} P\in I\!\!\!\,R^+\;\;\;\;\;\;\;\;\;\;\;\;\;\;\;\;\;\;\;\;\;
{\rm if} N\in\gamma_4\;.\;\;\;\;\;\;\;\;\;\;\;\;\;\;\;\;\;\;\;\end{equation}
\renewcommand{\theequation}{\arabic{equation}}
Therefore, the various acceptable choices of integration contour for the lapse
function in the path integral lead (like in the case of the scalar field
model with $\cosh$ potential) to wave functions in different
quantum theories, each of them associated with a different set of reality
conditions. If we start by imposing a set of reality conditions such that
the real $(w,z)$ representation is well defined, the ambiguity in the
selection of complex contours of integration disappears completely.

On the other hand, it is easy to check that the wave functions (62) do not
belong to any of the Hilbert spaces $L^2(I\!\!\!\,R\pm i\,\tilde{\epsilon})$,
$L^2(I\!\!\!\,R)$ or $L^2(I\!\!\!\,R^+)$. Hence, the propagation amplitudes
that one gets by computing the path integral between fixed real
geometrodynamic configurations cannot be considered proper
quantum states, since they all possess infinite norms. In particular,
for the scalar field model with exponential potential this means that the
wave functions selected by the no-boundary proposal, which are the propagation
amplitudes for fixed initial configuration $w'=-z'=k/2$ (that is,
$x'=y'=0$, see (54.a)) [10], cannot be accepted as quantum states. The
no-boundary proposal fails again in this model to pick out a normalizable
wave function.

{}From the analysis in Ref. [11], we also know the wave functions chosen by
the tunneling proposals of Linde [19] and of Vilenkin [20] in the
anisotropic models discussed in this work. According to Halliwell and
Louko, these are the propagation amplitudes obtained by integrating the
lapse function in (60) over the positive imaginary axis\footnote{i.e.,
$\gamma=\gamma_4$, and then $G$ is a Green function for the Wheeler-DeWitt
equation.} for fixed initial condition $w'=0$, and $z'$ restricted to be
positive (in some cases, $z'=0$ is also allowed [11]). Since these
propagation amplitudes correspond to wave functions of the form (62), we
conclude that the tunneling proposals do not select admissible quantum
states.

In addition to this, it is possible to prove that, for the Kantowski-Sachs
model ($k=1$), the classical solutions for fixed initial
conditions $w'=z'=0$ ($c'=b'=0$, see (54.c)) describe geometries that
close smoothly at initial Euclidean time with the local topology of
$\bar{B}^3\times S^1$, where $\bar{B}^3$ is the closed disc in $I\!\!\!\,R^3$.
Furthermore, the surface terms for the constant initial time section in the
Hilbert-Einstein action [11] vanish when $w'=z'=0$. Therefore, the
elimination of these boundary terms do not alter the path integral.
Moreover, the variational problem associated with the Hilbert-Einstein
action without initial boundary terms and fixed initial configuration
$w'=z'=0$ is always well posed. Following then the prescription of Halliwell
and Louko for the minisuperspace implementation of the no-boundary
proposal [11], the propagation amplitudes (60) with initial variables
$w'=z'=0$ (and $\gamma=\gamma_j$, $j=1,2,3$, given by (59)) turn out
to be no-boundary wave functions for the Kantowski-Sachs model with
cosmological constant. Once again, these no-boundary wave functions are
not admissible quantum states, since, from our previous discussion, they
all have infinite norms.

In this way, the most popular boundary conditions in quantum
cosmology,\linebreak namely, the tunneling proposals of Linde and of
Vilenkin and the no-boundary proposal of Hartle and Hawking, do not
succeed in selecting normalizable wave functions. It thus seems necessary
to modify or replace these boundary conditions in order to determine a
quantum state that can describe the evolution of the Universe.

\section {Equivalence of the Quantization with Different Sets of
Reality Conditions}

In this section, we will show that, for each of the models studied in this
work, the quantum theories that we have obtained by assuming different
sets of reality conditions are in fact equivalent. In particular, this
result will allow us to extract relevant physical predictions from quantum
theories constructed with other than Lorentzian reality conditions.

Although the discussion to follow can be generalized to all sets of
reality conditions of the form (14,15), or of the form (14) and (27) for
restricted domains of the variable $P$ in the
anisotropic minisuperspaces and the scalar field model with exponential
potential, we will concentrate our attention on two illustrative examples:
the quantization of the scalar field model with hyperbolic cosine potential
and reality conditions given either by equation (11) or by
\begin{equation}
\hat{Q}^{\star}=\hat{Q},\;\;\;\hat{P}^{\star}=\hat{P},\;\;\;\hat{H}^{\star}
=e^{i\frac{2\pi}{3}}\hat{H}+k\left(1-e^{i\frac{2\pi}{3}}\right),\;\;\;
\hat{T}^{\star}=e^{-i\frac{2\pi}{3}}\hat{T},\end{equation}
on one hand, and by equation (11) or by
\begin{equation} \hat{Q}^{\star}=-\hat{Q}\;,\;\;\;\hat{P}^{\star}=-\hat{P}\;
,\;\;\;\hat{H}^{\star}=\hat{H}\;,\;\;\;\hat{T}^{\star}=\hat{T}\;,
\end{equation}
on the other hand. Notice that conditions (64) coincide with the
$\star$-relations (42) and  (44.a) for $j=2$ in (42), and that equation (65)
corresponds to relations (42) and (44.b), for $j=1$.

The $\star$-algebras for reality conditions (11) and (64), whose
operators we will designate by the respective subindices 1 and 2, turn out
to be isomorphic under the transformation
\renewcommand{\theequation}{\arabic{equation}.a}
\begin{equation} I_2(\hat{Q}_1)=\hat{Q}_2\;,\;\;\;\;\;\;\;\;\;\;\;\;\;\;\;
\;\;\;\;\;\;\;\;\;I_2(\hat{P}_1)=\hat{P}_2\;,\end{equation}
\setcounter{equation}{65}
\renewcommand{\theequation}{\arabic{equation}.b}
\begin{equation}
I_2(\hat{H}_1)=e^{i\frac{\pi}{3}}\hat{H}_2+k\left(1-e^{i\frac{\pi}{3}}\right)
\;,\;\;\;\;\;\;I_2(\hat{T}_1)=e^{-i\frac{\pi}{3}}\hat{T}_2\;.\;\;\end{equation}
\renewcommand{\theequation}{\arabic{equation}}
The above isomorphism leaves invariant the Hamiltonian constraint (8), as
well as the equations of motion of the model provided that the lapse
functions $N_1$ and $N_2$ in the respective theories with reality
conditions (11) and (64) are related by
\begin{equation} N_2=e^{i\frac{\pi}{3}}N_1\;,\end{equation}
with $N_1\in I\!\!\!\,R$. A similar result can be reached for the sets
of $\star$-relations (11) and (65), for a isomorphism between
$\star$-algebras given by
\renewcommand{\theequation}{\arabic{equation}.a}
\begin{equation} I_3(\hat{Q}_1)=-i\hat{Q}_3\;,\;\;\;\;\;\;
I_3(\hat{P}_1)=i\hat{P}_3\;,\end{equation}
\setcounter{equation}{67}
\renewcommand{\theequation}{\arabic{equation}.b}
\begin{equation} I_3(\hat{H}_1)=\hat{H}_3\;,\;\;\;\;\;\;\;\;I_3(\hat{T}_1)
=\hat{T}_3\;,\end{equation}
\renewcommand{\theequation}{\arabic{equation}}
and lapse functions $N_1=N_3\in I\!\!\!\,R$. Here, we have used the
subindex 3 to denote operators and functions in the theory with reality
conditions of the form (65).

In the $(T,P)$ representation discussed in subsection 2.1.2, and with
the same kind of notation that has been introduced above, the quantum
states (10) of all the theories under consideration adopt the expression
\begin{equation} \Phi_l(T_l,P_l)=e^{-ikT_l}f_l(P_l)\;,\end{equation}
with $l=1,2,3$, $\,T_1,T_3,P_1,P_2\in I\!\!\!\,R$, $T_2\in e^{i\pi/3}
I\!\!\!\,R$, $P_3\in -i \,I\!\!\!\,R$, $\,f_1(P),f_2(P)\in L^2(I\!\!\!\,R)$
and $f_3(P)\in L^2(-i I\!\!\!\,R)$.

The Hilbert spaces of the quantum theories obtained with reality conditions
(11) and (64) are then clearly isomorphic under the mapping
\begin{equation} \Phi_2(T_2,P_2)\equiv h_2[\Phi_1](T_2,P_2)=\Phi_1(e^{-i
\frac{\pi}{3}}T_2,P_2)\;.\end{equation}
The same conclusion is applicable to the Hilbert spaces selected by
relations (11) and (65). In this case, the isomorphism is provided by
\begin{equation} \Phi_3(T_3,P_3)\equiv h_3[\Phi_1](T_3,P_3)=\Phi_1(T_3,
iP_3)\;.\end{equation}
It is not difficult to check that, in the adopted representation, the
isomorphisms (70) and (71) are compatible, respectively, with those
established before in (66) and (68) for the corresponding
$\star$-algebras.  We conclude then that the quantum theories constructed
with the sets of reality conditions (11), (64) and (65) result in being all
equivalent, as we had anticipated.

On the other hand, using the different isomorphisms introduced in this
section, it is now straightforward to obtain Lorentzian physical
predictions in the representations of the quantum theory that correspond
to the non-Lorentzian reality conditions (64) or (65). By way of an
example, let us consider the ``Lorentzian'' expectation value of the
configuration variable $y$ in the quantum state represented by the
different wave functions $\Phi_l$ ($l=1,2,3$) of the type (69), with
\begin{equation} f_1(P)=f_2(P)=f_3(-iP)\;\;\;{\rm and} \;\;\;P\in
I\!\!\!\,R \;.\end{equation}
Taking into account the definition of the operator $\hat{y}$ given in
(37), and equations (66) and (68), we arrive at the following
formulas in the representations $l=1$, 2 and 3:
\renewcommand{\theequation}{\arabic{equation}.a}
\begin{equation} <y>_1=\int_{I\!\!\!\,R}dP\;\overline{f_1}(P)
\left(i\frac{\partial f_1}{\partial P}(P)-8kTP\,f_1(P)\right)\;;
\;\;\;\;\;\; \end{equation}
\setcounter{equation}{72}
\renewcommand{\theequation}{\arabic{equation}.b}
\begin{equation} <y>_2=\int_{I\!\!\!\,R}dP\;\overline{f_2}(P)
\left(i\frac{\partial f_2}{\partial P}(P)-8ke^{-i\frac{\pi}{3}}T_2Pf_2(P)
\right),\;\;\;T_2\in e^{i\frac{\pi}{3}}I\!\!\!\,R\;;\end{equation}
\setcounter{equation}{72}
\renewcommand{\theequation}{\arabic{equation}.c}
\begin{equation} <y>_3=i\int_{(-i I\!\!\!\,R)}dP\; \overline{f_3(P)}
\left(\frac{\partial f_3}{\partial P}(P)-i8kTP\,f_3(P)\right)\;;\;\;\;\;\;\;
\end{equation}
\renewcommand{\theequation}{\arabic{equation}}
all of which coincide under the assumption (72). Finally, notice that the
expectation value $<y>$ is time dependent, as it should be for a
variable that is not an observable of the model.

\section {Conclusions}

By completing the nonperturbative quantization programme in three types of
minisuperspaces with different sets of reality conditions, and finding the
transformation that changes from the representation chosen in the
nonperturbative quantization to the metric representation used in the
path-integral approach, we have proved that the selection of complex
contours in the path integrals corresponds in fact to the choice of sets
of reality conditions for which the real metric representation can be
defined. Therefore, the ambiguity in the choice of complex contours of
integration disappears when one imposes an adequate set of reality
conditions.

We have also demonstrated that the propagation amplitudes obtained by
means of the complex path integral are, in general, non-normalizable
wave functions, and so they cannot be accepted as proper quantum states.
For the models studied here, this conclusion is also applicable to the
wave functions of the Universe determined by the path-integral
implementation of the no-boundary condition and the tunneling proposals,
up to date the most successful boundary conditions in quantum cosmology.

Finally, we have shown with some illustrative examples that different
sets of reality conditions can lead to equivalent quantum theories.
As a consequence, in some cases it is possible to gain physical
predictions corresponding to Lorentzian gravity from quantum theories
constructed with other than Lorentzian reality conditions. For more
complicated models than those considered in this work, the generalization
of this result would enable us to achieve the nonperturbative quantization
by finding a set of reality conditions that are equivalent to the
Lorentzian ones but simpler to impose in the quantization programme [21].
In this way, one could solve the technical difficulties that are usually
encountered in the nonperturbative quantum theory of gravitation when
implementing the reality conditions associated with Lorentzian gravity.

{\bf Acknowledgements}

\vspace*{.4cm}
The author is greatly thankful to A. Ashtekar, J. Louko, N. Manojlovi\'c
and D. Marolf for helpful discussions and valuable comments. He wants to
thank also the Departments of Physics at Syracuse University and the
Pennsylvania State University for warm hospitality. This work was
supported by funds provided by the Spanish Ministry of Education and
Science Grant No. EX92-06996911.


\newpage

\begin{thebibliography}{21}

\bibitem{1} Ashtekar A 1986 {\it Phys. Rev. Lett.} {\bf 57} 2244; 1987 {\it
Phys. Rev.} D {\bf 36} 1587

\bibitem{2} Ashtekar A 1991 {\it Lectures on Non-Perturbative Canonical
Gravity} ed Fang L Z and Ruffini R (Singapore: World Scientific)

\bibitem{3} Ashtekar A {\it The Proceedings of the 1992 Les Houches School on
Gravitation and Quantization} ed Julia B (Amsterdam: North Holland) to
be published

\bibitem{4} Bengtsson I 1988 {\it Class. Quantum Grav.} {\bf 5} L139; 1990
{\it Class. Quantum}\newline {\it Grav.} {\bf 7} 27

Kodama H 1988 {\it Prog. Thor. Phys.} {\bf 80} 1024; 1990 {\it Phys. Rev.}
D {\bf 42} 2548

Koshti S and Dadhich N 1989 {\it Class. Quantum Grav.} {\bf 6} L223

Ashtekar A and Pullin J 1990 {\it Ann. Israel Phys. Soc.} {\bf 9} 65

Bombelli L and Torrence R J 1990 {\it Class. Quantum Grav.} {\bf 7} 1747

Kastrup H A and Thiemann T 1993 {\it Nucl. Phys.} B {\bf 399} 211

Manojlovi\'c N and Mikovi\'c A 1993 {\it Class. Quantum Grav.} {\bf 10}
559

Manojlovi\'c N and Mena Marug\'an G A 1993  Nonperturbative canonical
quantization of minisuperspace models: Bianchi types I and II {\it Phys. Rev.}
D {\bf 48} to appear

\bibitem{5} Ashtekar A, Tate R and Uggla C 1993 Minisuperspaces: observables
and quantization {\it Preprint} Syracuse University

\bibitem{6} Halliwell J J 1990 {\it Proceedings of the
Jerusalem Winter School on Quantum Cosmology and Baby
Universes} ed Coleman S, Hartle J B and Piran T (Jerusalem)

\bibitem{7} Matzner R A and Mezzacapa 1986 {\it Found. Phys.} {\bf 16} 227

\bibitem{8} Halliwell J J 1987 {\it Phys. Lett.} B {\bf 185} 341

\bibitem{9} Lucchin F and Matarrese S 1985 {\it Phys. Rev.} D {\bf 32} 1316

Ratra B 1989 {\it Phys. Rev.} D {\bf 40} 3939

\bibitem{10} Garay L J, Halliwell J J and Mena Marug\'an G A 1991 {\it Phys.
Rev.} D {\bf 43} 2572

\bibitem{11} Halliwell J J and Louko J 1990 {\it Phys. Rev.} D {\bf 42} 3997

\bibitem{12} Ashtekar A, Romano J D and Tate R S 1989 {\it Phys. Rev.} D {\bf
40} 2572

\bibitem{13} Rendall A 1993 Unique determination of an inner product by
adjointness\newline  relations in the algebra of quantum observables {\it
Preprint} Syracuse University SU-GP-93/2-2

\bibitem{14} Halliwell J J 1988 {\it Phys. Rev.} D {\bf 38} 2468

\bibitem{15} Halliwell J J and Hartle J B 1991 {\it Phys. Rev.} D {\bf 43}
1170

\bibitem{16} Halliwell J J and Hartle J B 1990 {\it Phys. Rev.} D {\bf 41}
1815

\bibitem{17} Abramowitz M and Stegun I A (ed) 1965 {\it Handbook of
Mathematical Functions} (Natl. Bur. Stand. Appl. Math. Ser. No. 55)
(Washington, D.C.: U.S. Govt. Print. Off.)

\bibitem{18} Hawking S W 1982 {\it Astrophysical Cosmology} ed Br\"uck H A,
Coyne G V and Longair M S (Vatican City: Pontificia Academia Scientarium);
1984 {\it Nucl. Phys.} B {\bf 239} 257

Hartle J B and Hawking S W 1983 {\it Phys. Rev.} D {\bf 28} 2960

\bibitem{19} Linde A 1984 {\it Zh. Eksp. Teor. Fiz.} {\bf 87} 369 (1984 {\it
Sov. Phys. JETP} {\bf 60} 211); 1984 {\it Nuovo Cimento} {\bf 39} 401;
1984 {\it Rep. Prog. Phys.} {\bf 47} 925

\bibitem{20} Vilenkin A 1984 {\it Phys. Rev.} D {\bf 30} 509; 1986 {\it Phys.
Rev.} D {\bf 33} 3560; 1988 {\it Phys. Rev.} D {\bf 37} 888

\bibitem{21} Mena Marug\'an G A 1993 in preparation

\end{thebibliography}
\end{document}